# Advanced EMT and Phasor-Domain Hybrid Simulation with Simulation Mode Switching Capability for Transmission and Distribution Systems

Qiuhua Huang, *Member, IEEE*, and Vijay Vittal, *Fellow, IEEE*

*Abstract*—Conventional electromagnetic transient (EMT) and phasor-domain hybrid simulation approaches presently exist for transmission system level studies. Their simulation efficiency is generally constrained by the EMT simulation. With an increasing number of distributed energy resources and non-conventional loads being installed in distribution systems, it is imperative to extend the hybrid simulation application to include distribution systems and integrated transmission and distribution systems. Meanwhile, it is equally important to improve the simulation efficiency as the modeling scope and complexity of the detailed system in the EMT simulation increases. To meet both requirements, this paper introduces an advanced EMT and phasor-domain hybrid simulation approach. This approach has two main features: 1) a comprehensive phasor-domain modeling framework which supports positive-sequence, three-sequence, three-phase and mixed three-sequence/three-phase representations and 2) a robust and flexible simulation mode switching scheme. The developed scheme enables simulation switching from hybrid simulation mode back to pure phasor-domain dynamic simulation mode to achieve significantly improved simulation efficiency. The proposed method has been tested on integrated transmission and distribution systems. The results show that with the developed simulation switching feature, the total computational time is significantly reduced compared to running the hybrid simulation for the whole simulation period, while maintaining good simulation accuracy.

*Index Terms*-- EMT and phasor-domain hybrid simulation, multi-area Thévenin equivalent, simulation mode switching.

## I. Introduction

POWER system dynamic analysis requirements are significantly evolving due to the changing nature of generation and loads. A significant portion of newly installed generation resources and a variety of loads including induction motors are interfaced with the grid through power electronic (PE) converters [1]. However, existing transient stability (TS) simulation tools cannot adequately represent PE devices in sufficient detail, particularly during the faulted period [2]-[7]. Additionally, these tools cannot adequately represent single-phase air-conditioner compressor motors [9], which account for an increasingly large percentage of summer peak load in many power systems [10]. These modeling inadequacies can result in either an over- or under-estimate of the system reliable operation boundary and/or stability limits, which in turn results in systems operating under either higher risk or less efficient conditions. On the other hand, EMT simulation tools can simulate PE and single-phase devices in detail. However, they are not suitable for large system-level studies. Several different simulation methods, including EMT-TS hybrid simulation [2]-[8], frequency adaptive simulation [11] and dynamic phasor based approach [12], were proposed or developed to solve similar simulation challenges. Among them, the hybrid simulation approach has attracted the most attention from both industry and academia.

Previous research and development efforts on hybrid simulation mainly focused on the following three aspects: 1) network equivalent, including fundamental frequency based equivalent [6]-[8] and frequency dependent network equivalents [13]; 2) interaction protocols for improving efficiency and performance during fault periods [6]-[8]; 3) hybrid simulation program development [13]-[16]. Except for [15], others were developed only for transmission systems.

Except for the real-time simulator-based hybrid simulation, the overall computational efficiency of hybrid simulation for realistically large system applications is still not satisfactory, as shown in a recent study in [8]. It should be noted that the computational efficiency issue was not emphasized in previous studies mainly because the EMT (or detailed) systems in these studies were relatively small and simple [2]-[3], [5]-[7]. Given the present trend of increasing application of non-conventional generation and loads as well as HVDC transmission [1], it is expected that a larger portion of the system with more complex and detailed models needs to be modeled in the EMT simulator. Consequently, the computational efficiency issue of the EMT simulator will constitute a significant bottleneck. While the performance of EMT simulators are improving, the improvement is not sufficient compared to the ever-increasing computational demands. One alternative to improve the computational efficiency of hybrid simulation is to enhance the simulation workflow. Previous studies in [2] and [3] showed that the overall simulation efficiency could be significantly

This work was supported by the National Science Foundation under the Grant EEC-9908690 at the Power System Engineering Research Center.
Q. Huang is with the Electricity Infrastructure Group, Pacific Northwest National Laboratory, Richland, WA, 99354 USA (e-mail: qiuhua.huang@pnnl.gov ).
V. Vittal is with the Department of Electrical, Computer and Energy Engineering, Arizona State University, Tempe, AZ 85287-5706 USA (e-mail: vijay.vittal@asu.edu ).

improved by transitioning from the hybrid simulation mode back to TS simulation mode after the fast dynamics settle down. However, there are some outstanding limitations in the underlying assumptions and implementation in [2] and [3]. More detailed discussions will be presented in Section II.

Furthermore, almost all the previous hybrid simulation research and development efforts focused on transmission systems only, and very few on distribution systems until recently [15], and none on integrated transmission and distribution (T&D) systems. With an increasing number of distributed energy resources (DER) and new types of loads being installed in the distribution systems, it is imperative to extend the hybrid simulation application to distribution systems [15] and integrated T&D systems.

In this paper, an advanced hybrid simulation program has been developed by extending the hybrid simulation program developed by the authors in [8],[14]. Compared to the earlier work in [8],[14], the new hybrid simulation program not only has comprehensive phasor-domain modeling capabilities, supporting three-phase, three-sequence and conventional positive-sequence modeling, but also features a robust ability to switch from the hybrid simulation mode back to phasor-domain dynamic simulation (DS) mode. The main contributions of the paper include: 1) a comprehensive hybrid simulation program suitable for transmission, distribution as well as integrated T&D systems, instead of separate hybrid simulation programs for each domain; 2) a robust simulation mode switching scheme. To realize simulation mode switching, discrete events and/or control signals captured by the EMT simulation are utilized to reconcile the simulation results of the phasor-domain and EMT representations of the detailed system.

The remainder of the paper is organized as follows: A review of hybrid simulation with simulation mode switching is presented in Section II. Comprehensive phasor-domain modeling and dynamic simulation capabilities developed for hybrid simulation are discussed in Section III. In Section IV, development of an advanced hybrid simulation program is presented in detail. Test cases and results are shown in Section V. Additional discussions and conclusions are provided in Sections VI and VII, respectively.

## II. HYBRID SIMULATION WITH SIMULATION MODE SWITCHING

### A. The Concept of Simulation Mode Switching

Conventional hybrid simulation consists of two stages, as illustrated in Fig. 1. It starts with the TS simulation stage, then switches to the hybrid simulation stage prior to the fault occurrence and remains in the mode till the end of simulation.

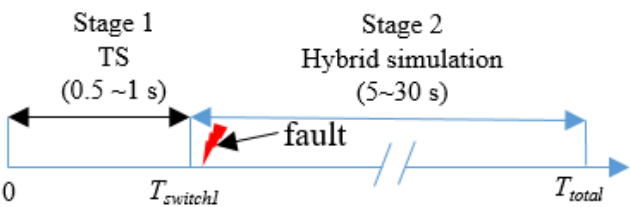

Fig. 1 A schematic illustrating conventional 2-stage hybrid simulation

The primary motivation behind switching from TS simulation to EMT-TS hybrid simulation is to take advantage of the detailed modeling and simulation capabilities of the EMT simulation. Fast dynamics in power systems usually settle down in a short time period after the fault is cleared. When the dominant dynamics in the system are in the low frequency range, these components modeled in the EMT simulator can be represented by their corresponding phasor-domain dynamic or quasi-steady-state (QSS) models. Therefore, it is theoretically feasible to switch from EMT-TS hybrid simulation back to pure TS simulation after the fast dynamics settle down without significantly compromising the simulation accuracy.

The two-stage hybrid simulation (depicted in Fig. 1) provides users the "zoom-in" capability. When the conventional hybrid simulation is enhanced with the simulation mode switching capability, as shown in Fig. 2, users can "zoom-out" from the detailed modeling and simulation by transitioning from hybrid simulation back to phasor-domain dynamic simulation after the fast dynamics in the detailed system settle down. Such a "zoom-in/zoom-out" switching capability not only provides more flexibility to the users, but also reduces the simulation time significantly for hybrid simulation applications that involve a relatively large and/or complex detailed system. It should be noted that the durations shown in the Figs. 1 and 2 are intended to give users the basic idea regarding how long each stage usually lasts in practical applications, thus they should not be treated as exact numbers. The exact numbers vary from case to case.

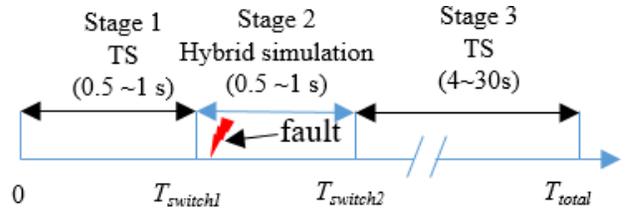

Fig. 2. Advanced hybrid simulation with simulation mode switching

Switching from stage 1 (TS) to stage 2 (hybrid simulation) is relatively easy to achieve, since it is completed under the pre-fault, normal steady state operating condition. In contrast, switching from stage 2 (hybrid simulation) to stage 3 (TS) is more complicated and challenging since the system is operating under a non-equilibrium, dynamic condition.

### B. Previous Approaches and Key Issues

In the approach proposed in [2], the detailed system has two representations, i.e., an EMT representation and a phasor model representation. During the stage 2 shown in Fig. 2, the phasor model representation of the detailed system is simulated using TS simulation in parallel with the hybrid simulation. The switching operation is performed when the simulation results of both representations converge. A fundamental issue with this approach lies in the implicit assumption that simulation results of the two representations of the detailed system would converge at some point after the fast dynamics settle down. However, the convergence is not guaranteed without reconciling the simulation results, due to the inherent differ-ences in the modeling and simulation methods between EMT and TS simulations. Furthermore, the implementation was only discussed at a high level, with many important implementation details not provided in [2].

Another simulation switching approach proposed in [3] directly initializes a phasor model representation of the detailed system before switching back to the TS simulation. It was achieved by simply assuming the states of the detailed system would fully recover to the pre-fault steady-state condition, which obviously is too strong an assumption and thus unrealistic for many hybrid simulation applications.

## III. COMPREHENSIVE PHASOR-DOMAIN MODELING AND SIMULATION CAPABILITIES

### A. Comprehensive Phasor-Domain Modeling Capabilities

Conventional transient stability simulation is developed based on positive-sequence phasor modeling of all power system components. Most of previous hybrid simulation solutions were developed based on positive-sequence phasor modeling [2]-[3],[6]-[7]. However, such a modeling approach is only suitable for three-phase balanced conditions. Consequently, when there are unbalanced conditions in the detailed system and/or the external system, it is challenging to interface with an EMT simulation. To overcome this drawback, the authors have developed a three-sequence transient stability simulation algorithm for an improved hybrid simulation algorithm in [8]. For both positive-sequence and three-sequence modeling approaches, the resulting hybrid simulation is only suitable for modeling transmission systems in the external system. One important objective of this paper is to further enhance the modeling capabilities of the program in [8] such that the new hybrid simulation program can be readily applied to integrated T&D systems and distribution systems. This means the new program should support the three-phase modeling capability for the distribution systems. A three-phase and mixed three-phase/three-sequence modeling framework has been developed in the authors' previous work [17]. Thus, the phasor-domain dynamic simulation tool used in the paper has comprehensive modeling capabilities as shown in Fig. 3.

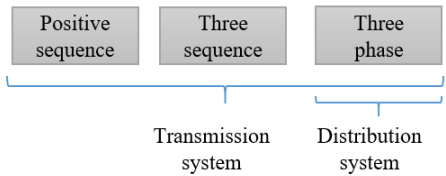

Fig. 3. Comprehensive phasor modeling capabilities are developed for transmission, distribution and integrated T&D systems

### B. Multi-Area Thévenin Equivalent based Dynamic Simulation

A multi-area Thévenin equivalent (MATE) based dynamic simulation algorithm has been developed in [17]. With this simulation approach, the whole system can be partitioned into several subsystems. The subsystems are interconnected through link branches (as shown in Fig. 3 in [17]). The MATE approach is used to reconcile the individual solutions of subsystems into a full-system simultaneous solution at the network solution stage for each time step. More details of the MATE-based dynamic simulation can be found in [17].

## IV. PROPOSED HYBRID SIMULATION WITH AN IMPROVED SIMULATION MODE SWITCHING SCHEME

### A. Overview of the Proposed Hybrid Simulation Design

The proposed hybrid simulation solution for switching from the hybrid simulation to the TS simulation extends the basic idea of the approach discussed in [2]. In the proposed approach, the hybrid simulation program developed by the authors in [8], the comprehensive modeling framework and the MATE-based dynamic algorithm presented in Section III are integrated. In this research, PSCAD/EMTDC [21] is selected as the EMT simulator and the open source power system simulation engine InterPSS [22] is chosen as the phasor-domain simulator. Significant improvements have been made to address the drawbacks of the original approach proposed in [2]. The main issue of discrepancy in simulation results between the EMT and TS simulation of the detailed system is addressed from both the modeling and simulation result coordination perspectives:

Firstly, from the modeling perspective, the three-phase phasor modeling inherently relates more closely to the three-phase point-on-wave modeling used in the EMT simulation than the positive-sequence and three-sequence phasor modeling. The developed program supports three-phase phasor modeling. This feature is particularly useful when there are some models (e.g., single-phase induction motors) and or contingencies that are better modeled in phase-oriented representation in the detailed system.

Secondly, it is observed from past simulation experiences that major simulation result discrepancies between the EMT and the TS simulations at the post-disturbance stage are usually related to some control actions and operation state changes of some critical components modeled in the detailed system representation. The main reason for this discrepancy is that these actions and/or state changes may not be correctly represented by the phasor representation or captured by the TS simulation due to some inherent modeling and simulation limitations [18]. Therefore, in the proposed approach, besides the updated network equivalent data, these discrete event or control signals obtained from EMT simulation results are transferred back to the TS simulation of the detailed system. These signals are used as external control inputs to override the corresponding control signals or to change the states of the corresponding models in the TS simulation. These important signals can be, for example, a converter blocking signal when an HVDC inverter is undergoing commutation failure or a state change signal of an A/C motor when the motor stalls.

In addition, re-connecting the two systems into a full network becomes unnecessary with the proposed approach. After switching back to the TS simulation mode, the detailed and the external systems are still modeled as two subsystems in the MATE-based dynamic simulation.

With the proposed simulation algorithm, the full system is split into two parts, i.e., the detailed system and the external system, using the bus splitting method as shown in Fig. 4(a)-(b). For each boundary bus, a dummy bus is created during the network splitting stage. In order to make the splitting scheme compatible with the MATE-based dynamic simulation algorithm, a "virtual breaker" is introduced to link the original boundary bus and the dummy bus. The detailed system has two representations, i.e., EMT- and phasor-domain representations, as shown in Fig. 4(c).

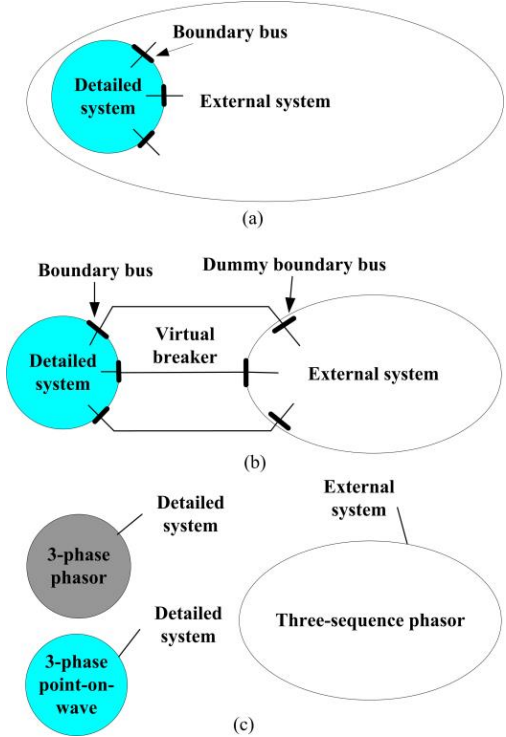

Fig. 4 (a) the full network (b) the full network is split into the detailed system and the external system connected by virtual breakers; (c) representations of the detailed system and the external system used in the proposed method

The EMT representation of the detailed system is developed based on the requirement of the phenomenon of interest and detailed models involved in the study. The phasor-domain representation of the detailed system is developed with reference to the EMT representation counterpart, by substituting the detailed models in EMT representation with corresponding phasor domain models, to ensure both representations are as consistent as possible. Models of both representations of the detailed system and the external system have to be developed before hybrid simulation. The external system is by default modeled in the three-sequence phasor representation.

The interactions between the detailed system and the external system for the three stages are shown in Fig. 5. The mutual representations of the detailed and external systems are different at different stages of hybrid simulation. As shown in Fig. 5(a), at the stages 1 and 3 of hybrid simulation, the Thévenin equivalents of the detailed and external systems are calculated for each system. But they are not directly added to each system, instead they are used to build a link subsystem, which produces the current injections at the boundary for the detailed and external systems. This is the multi-area Thévenin equivalent approach [17] presented in Section III.B.

As shown in Fig. 5(b), at the stage 2 of hybrid simulation, the detailed system is represented as three-sequence current injections (extracted from the instantaneous waveforms) in the external system, while the external system is represented as three-phase Thévenin equivalent. The approach shown in Fig. 5(b) will be discussed in detail in the following subsection B.

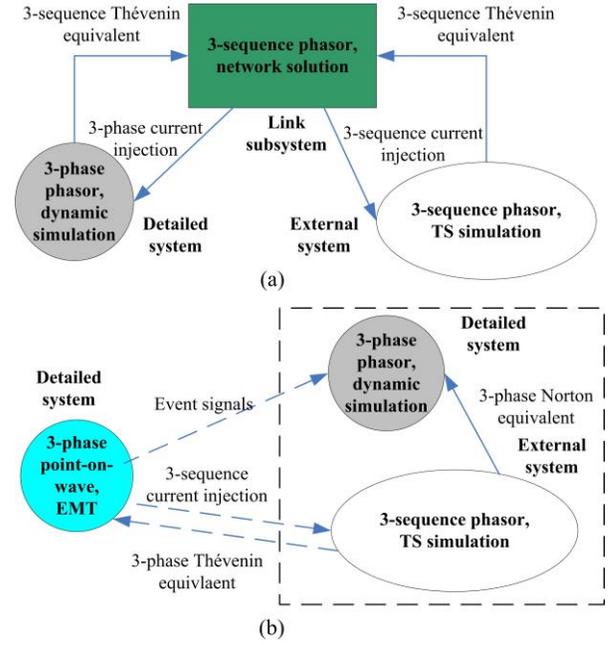

Fig. 5. Interactions and equivalent data exchanges between the detailed and the external systems: (a) for both stages 1 and 3; (b) for stage 2

### B. Extend Conventional Hybrid Simulation Program to Realize Simulation Mode Switching

For the second stage, the hybrid simulation program developed in [8] has been extended. The hybrid simulation is augmented by running the dynamic simulation of the detailed system and the switch controller in parallel to the existing hybrid simulation algorithm, as shown in Fig. 5(b). The actual implementation of the augmented hybrid simulation algorithm is illustrated in Fig. 6. The enhancement to the original development in [8] is shown within the dashed rectangular in Fig. 6(a). In both figures, $t$ denotes the start time for the processing step, $\Delta T$ is TS simulation time step as well as the EMT-TS interaction time step, $I_{EMT(t)}^{120}$ and $I_{EMT(t-\Delta T)}^{120}$ are the three-sequence current injection vectors sent from the EMT side at the present and previous interaction time step, respectively; $x_{de(t)}$ and $x_{ex(t)}$ denote the monitored state variables of the detailed and external systems; $V_{de(t)}^{abc}$ denotes three-phase voltages of the boundary bus of the phasor-domain detailed system; $V_{ex(t)}^{120}$ represents three-sequence voltages of the boundary buses of the external system.

The sequence of the main steps at stage 2 of hybrid simulation is also shown in the circles in Fig. 6(a). The first step is data transfer from EMT side to the phasor domain simulator via a socket and pre-processing. Then, in step ②, $I_{EMT(t)}^{120}$ is used as the input for the three-sequence TS simulation and the three-sequence voltages of the boundary buses $V_{ex(t+\Delta T)}^{120}$ are updated. Subsequently, the three-phase Thévenin equivalent

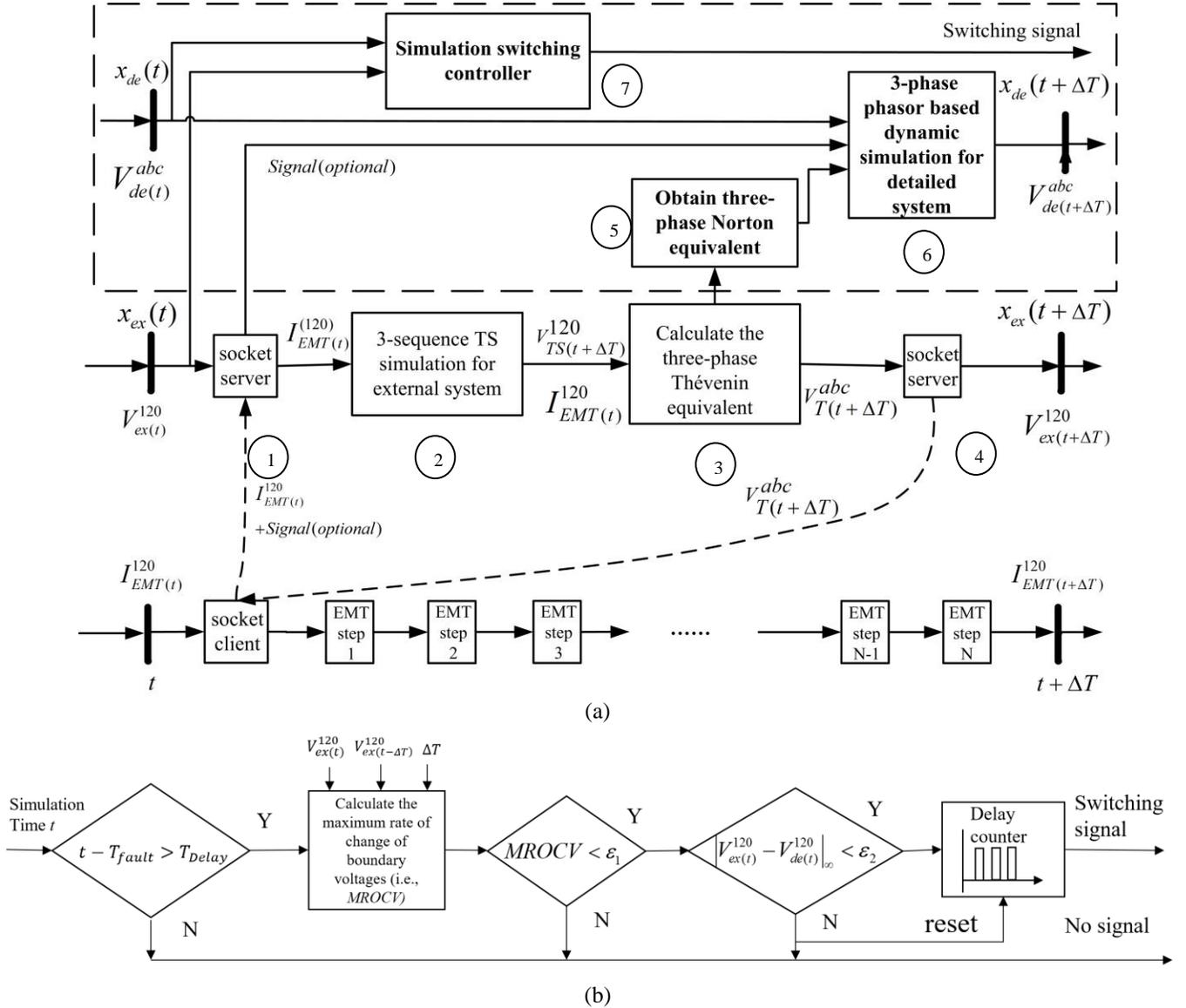

Fig. 6. Implementations of the augmented hybrid simulation at the stage-2: (a) the overall simulation process (modified based on the Fig. 2 in [8]); (b) the logic of the simulation switching controller in step ⑦

voltages $V_{T(t+\Delta T)}^{abc}$ are derived in step ③ and sent back to the EMT side in step ④. On receiving $V_{T(t+\Delta T)}^{abc}$, the EMT simulator continues to run consecutive multi-step EMT simulations in parallel of phasor-domain simulation until the next interaction time step. At the same time, a three-phase Norton equivalent is converted from the three-phase Thévenin equivalent in step ⑤, and used as the external system equivalent in the dynamic simulation of the detailed system in step ⑥. In the last step ⑦, the hybrid simulation switching controller determines whether the simulation mode can be switched from hybrid simulation to pure phasor domain simulation at the next time step.

The main reason for using Norton equivalent is that the network solution is usually formatted as $I(x,V) = YV$ in the dynamic simulation, and the form of Norton equivalent fits the formulation very well, with the current source part being added to the left-hand side, and the admittance part being added to the right-hand side. The admittance matrix of the detailed system is augmented with admittance matrix of the three-phase Norton equivalent after switching from the stage 1 to stage 2. The three-phase current source of the Norton equivalent is directly used in the network solution step. When switching from stage 2 to stage 3, the admittance matrix of the detailed system has to be rebuilt to account for the removal of the admittance matrix of the three-phase Norton equivalent.

The simulation mode switching process is controlled by the hybrid simulation switching controller. There are two basic principles behind this controller: 1) simulation switching is only performed after the system enters a slow dynamic state; 2) simulation switching is only performed after the boundary conditions of the detailed and external systems truly converge. Based on these two principles, this controller is designed as shown in Fig. 6(b). It includes three main parts, i.e., a time delay to bypass the transient period, system state monitoring based on the maximum rate of change of boundary voltages, checking the convergence of results of the two models of the detailed system. Firstly, a time delay is necessary to account for the period of the system transiting from a fast dynamic (transient) state to a slow dynamic state where the system can be adequately represented by phasor models. Based on the simulation experiences, the delay setting ($T_{Delay}$) is chosen to be 0.2 s in this paper. Once the time delay criterion is met after a fault is cleared, the switching controller begins to check whether the maximum rate of change of the boundary voltages is small enough ($\varepsilon_1$ =0.005 pu in this paper) to ensure the systems truly enter the slow dynamic state. Secondly, the voltages of the boundary buses between the detailed system and the external system are monitored and the maximum difference between them is used as the indicator for simulation switching. At each step, this indicator is calculated and compared with a preset tolerance (for example, $\varepsilon_2$ =0.005 pu). Furthermore, in order to make sure the boundary conditions of the detailed and external systems truly converge, the maximum voltage difference must be within the preset tolerance for a reasonably long period (e.g., 3-5 cycles). This is implemented in the delay counter block at the end of Fig. 6(b). If both the mismatch and time period criteria are satisfied, the simulation switch controller outputs a signal to indicate that the simulation can be switched to the TS simulation starting from the following time step. With this implementation, the first two parts ensure the first principle is satisfied and the last part guarantees that two parts of the system converge before simulation mode switching.

## V. TEST CASE AND RESULTS

### A. Test case

The developed advanced hybrid simulation program has been tested with a modified IEEE 9 bus system [19]. In this test case, the original aggregated load at bus 5 is replaced by an equivalent sub-transmission and distribution system, as shown in Fig. 7. Modeling of the sub-transmission and distribution system is shown in Fig. 8. The loads are connected to an equivalent feeder. As for the load composition, 50% of the total load in terms of active power is represented by single-phase residential A/C motors, while the remainder is modeled as constant impedances. The portion of the system highlighted in Fig.7 and shown in Fig. 8 is modeled as the detailed system. The remainder of the system is modeled as the external system and represented in three-sequence phasor domain. As discussed in Section IV.A, there are two representations of the detailed system, i.e., EMT and phasor-domain. When the detailed system is simulated by EMT simulation, an EMT detailed A/C motor model developed in [9] is used. The important effect of the point-on-wave where a fault or voltage sag is initiated on motor stalling can be captured by this model [9]. On the other hand, when the detailed system is represented by phasor-domain and simulated by three-phase dynamic simulation, a performance model of A/C motor [20] is used to represent the A/Cs. The performance model represents the operation of the A/C motors with two states, i.e., running and stalled, and each state has its own performance curve. In this test case, the TS simulation time step is 0.005 s and the EMT simulation time step is 20 μs.

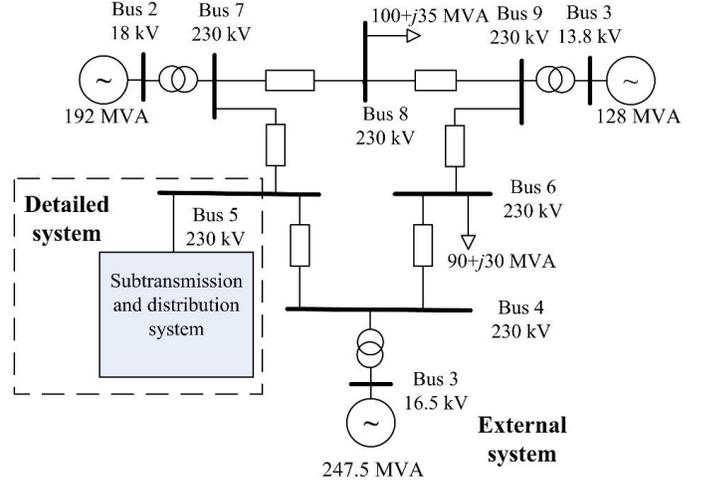

Fig. 7. One-line diagram of a modified IEEE 9 bus system

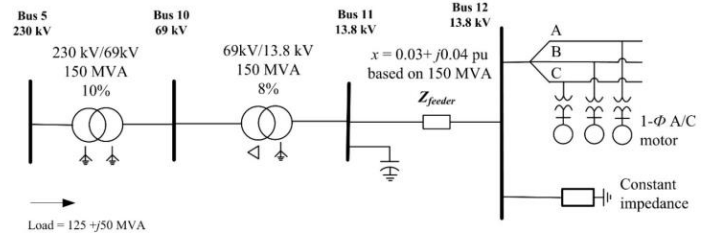

Fig. 8. The sub-transmission and distribution system served by bus 5

### B. Use of Discrete Event Signals to Reconcile the Two Detailed System Simulations

The following two simulation cases in this section are to study the convergence of results obtained from the EMT simulation and the 3-phase dynamic simulation for the detailed system during stage 2. The convergence of both results is the prerequisite for realizing simulation mode switching from hybrid simulation to pure phasor-domain simulation.

A single-line-to-ground (SLG) fault is applied to bus 10 at 0.5 s and cleared after 0.07 s. In order to study the effects of using discrete event signals from the EMT side to enhance the accuracy of TS simulation, the simulation switching function is intentionally disabled. The simulation results *without* and *with* sending the A/C motor operation status signals from the EMT simulation to the 3-phase dynamic simulation are shown in Fig. 9 and Fig. 10, respectively. A/C motor operation status signals indicate whether the A/C motors are running or stalled, which are used to determine the state of the corresponding A/C models in the phasor-domain representation. Note that both curves in Figs. 9 and 10 are obtained during the stage 2 of hybrid simulation for the two models of the detailed system. The curves labeled "EMT" are results of the EMT model (not full EMT simulation of the whole system), while the curves labeled "3-$\phi$ DS" are results of the phasor-domain model.

In the scenario without sending the A/C motor operation status signals from the EMT side to three-phase dynamic simulation, simulation results of the EMT and the dynamic simulation are significantly different in terms of A/C motor stalling. A/C motors on two phases stalled (status = 0) in the detailed system simulated by the 3-phase dynamic simulation. In contrast, the EMT simulation results show that only the A/C motor on phase *C* stalled. This difference is mainly due to the fact that the fault point-on-wave effects on A/C motor stalling [8],[9] cannot be accurately captured by the phasor-domain modeling and simulation. This difference results in significant differences between the two simulation results in the three-phase voltages of the boundary bus (bus 5 in this case), as shown in Fig. 9. Consequently, the proposed simulation switching criteria are never satisfied and the simulation keeps running in the hybrid simulation mode till the end of the whole simulation.

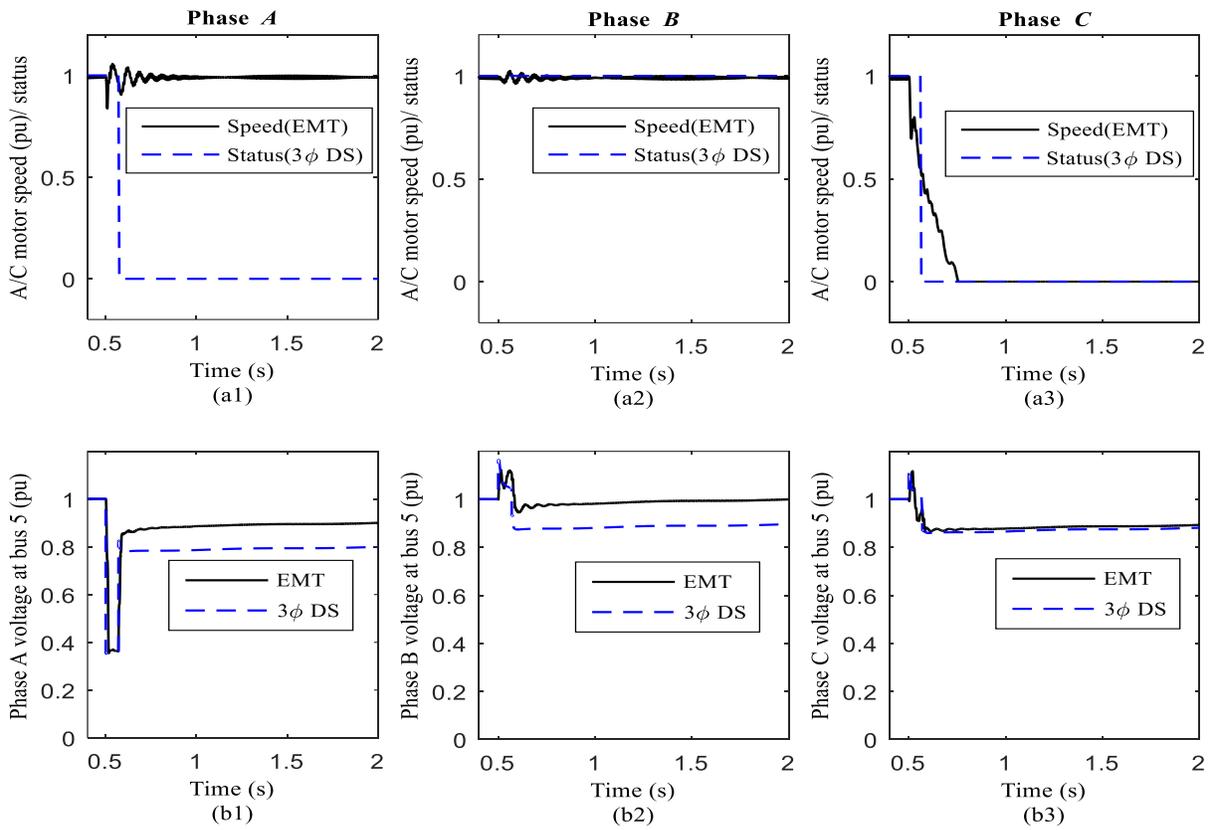

Fig. 9. Simulation results of the detailed system *without* sending the A/C motor status signals from the EMT simulation to the 3-phase dynamic simulation

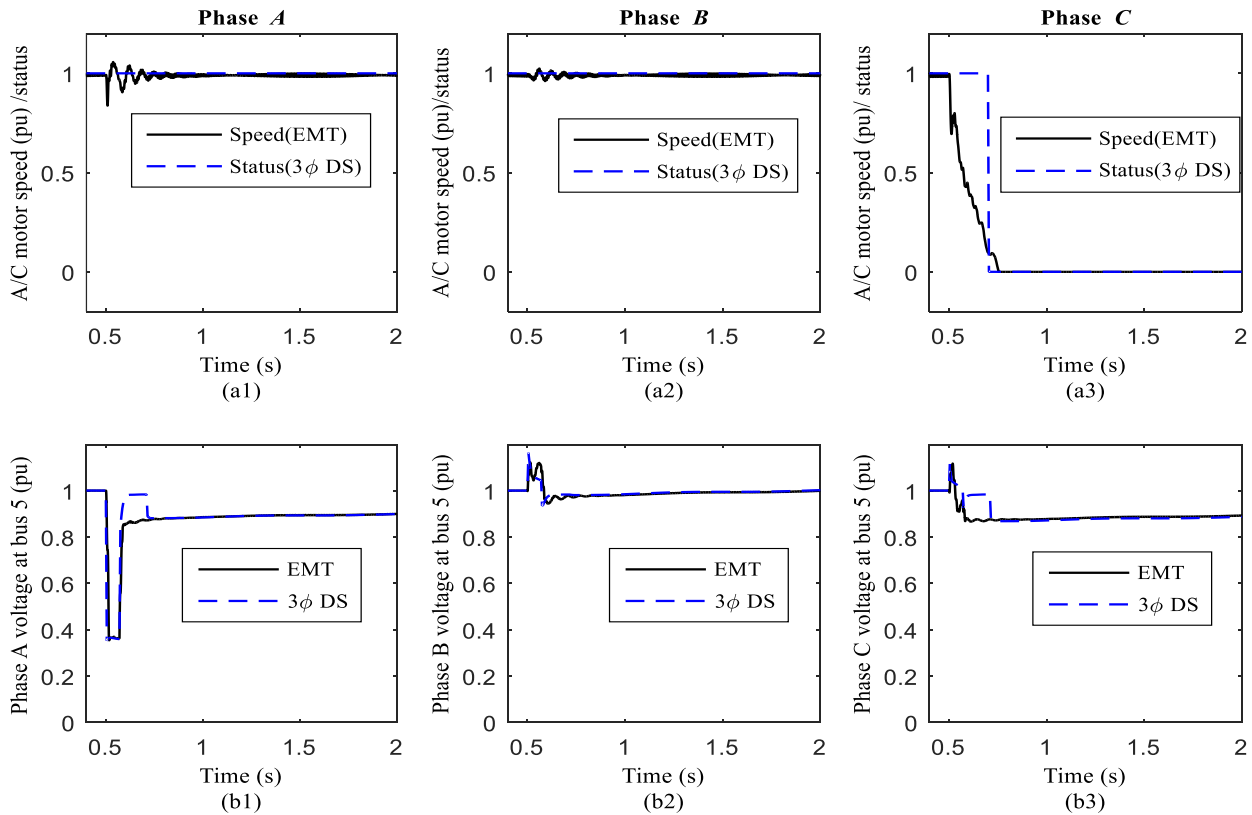

Fig. 10. Simulation results of the detailed system with sending the A/C motor status signals from the EMT simulation to the 3-phase dynamic simulation

In the scenario in which the A/C motor operation status signals are sent from the EMT side to the 3-phase dynamic simulation, the A/C stalling results are coordinated to make sure A/C motors in the three-phase dynamic simulation have the same status as those simulated by the EMT simulation during stage 2. While there are some discrepancies observed in the three-phase boundary bus voltages after the fault is cleared, the discrepancies last only a short time period (approximately 0.15 s) and disappear after the A/C motor on phase *C* effectively stalls in both the EMT simulation and the three-phase dynamic simulation.

The discrepancies in Fig. 10 discussed above are related to the A/C motor stalling process, which is adequately simulated in the EMT simulator, but cannot be represented by the performance model of the A/C motor in the dynamic simulation. During the A/C motor stalling process, the reactive power drawn by the A/C motor on phase *C* increases significantly when the A/C motor speed decreases to lower than 0.5 pu. The increased reactive power consumption depresses the voltages of both phase *A* and *C* of bus 5 after the fault is cleared, as shown in the solid black traces in Fig. 10(b1)-(b3). On the other hand, before receiving the A/C motor stalling signals from the EMT side, all three A/C motors operate in the running mode (status =1) in the detailed system simulated by the three-phase dynamic simulation. After the fault is cleared, the bus voltages recover to the levels close to their pre-fault values, as illustrated by the dashed blue traces in Fig. 10(b1)-(b3). These different responses of the A/C motors in the two detailed system simulations contribute to the discrepancies shown in Fig. 10.

### C. Criteria for switching from hybrid simulation to phasor-domain dynamic simulation

In this test, the switching criterion is as follows: starting at 0.2 s after the fault is cleared, if the maximum difference of boundary bus voltages (*maxΔV*) obtained by the dynamic simulation of the detailed and the external systems is less than 0.005 pu for more than 2 cycles, simulation will be switched from the hybrid simulation mode to the phasor-domain DS mode.

The results in the previous subsection showed that the proposed approach can significantly reduce the simulation result discrepancy caused by the modeling difference of A/C motors in the EMT and the TS simulation by coordinating the discrete motor stalling events in the EMT simulation with the TS simulation. With this approach applied to the test case, the monitored maximum bus voltage difference during stage 2 is shown in Fig. 11. The simulation is switched from the hybrid simulation to the pure TS simulation at 0.805 s, which is 0.235 s after the fault is cleared. It should be emphasized that the switching is performed under a dynamic, non-steady-state system condition, which is shown in the generator speeds in Fig. 12.

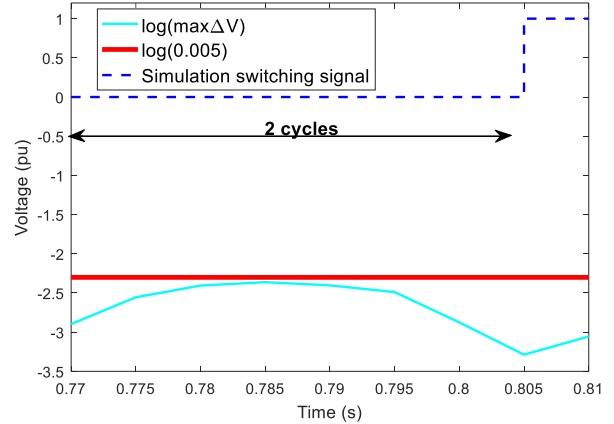

Fig. 11 The maximum voltage difference of boundary buses and the switching signal

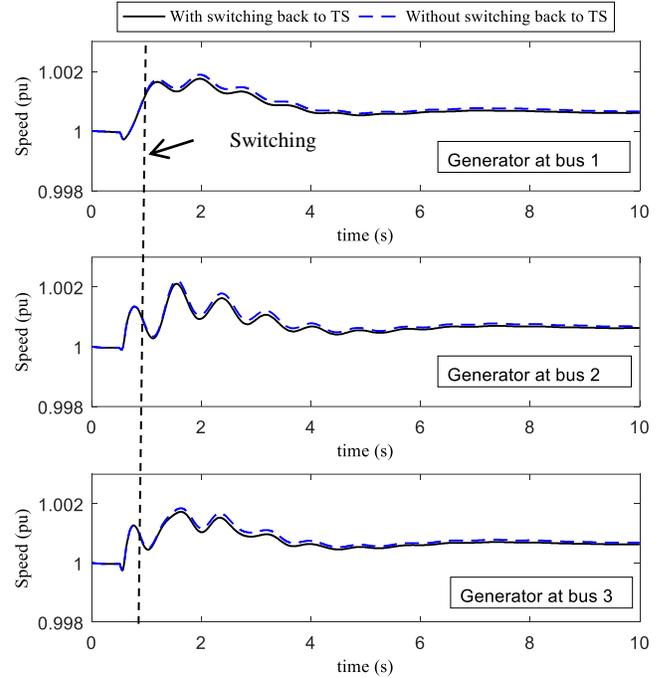

Fig. 12 Speeds of the generators at buses 1, 2 and 3

### D. Result comparisons

The simulation results are compared with those simulated by the conventional hybrid simulation approach without switching back to TS simulation, which are shown in Figs.12-13. The results of both approaches match closely, demonstrating the effectiveness of the proposed approach. The computational time for a 10-second simulation with different methods is shown in Table 1. Compared to the hybrid simulation without switching back to TS simulation, the developed hybrid simulation switching method reduces the computational time by 91.86 %.

Additionally, the developed program has been compared with conventional positive sequence TS simulation and full EMT simulation. The distribution system is modeled by single-phase representation (one A/C motor performance model for representing A/C motors on three phases) and the entire system is simulated positive sequence TS simulation. The results show that all A/C motors served by the substation of bus 5 stall for the same SLG fault at bus 10. This means conventional TS simulation produces significantly different results from that simulated by hybrid simulation. To illustrate the differences, the positive sequence voltages of bus 5 in both results are shown in Fig. 14.

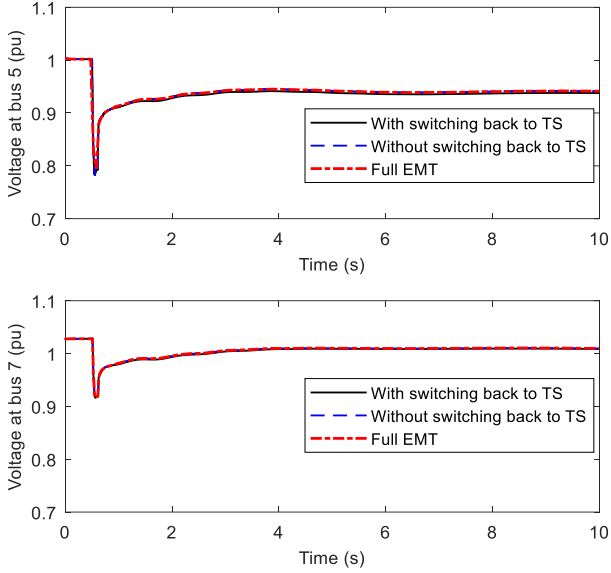

Fig. 13 Positive sequence voltages of bus 5 and bus 7 simulated by hybrid simulations and full EMT simulation.

Table I
The computational time of different methods

| Simulation method | Computational time /s |
| --- | --- |
| Hybrid simulation *without* switching back to TS simulation (developed in [8], [14]) | 189.2 |
| Hybrid simulation *with* switching back to TS simulation (developed in this paper) | 15.4 |

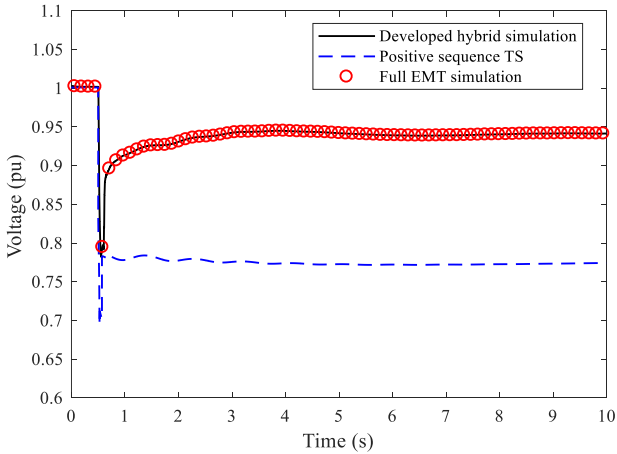

Fig. 14 Positive sequence voltages of bus 5 simulated by positive sequence TS simulation, full EMT simulation and the developed hybrid simulation (the full EMT simulation results has been re-sampled at a larger time step for this plot)

## VI. Discussions

While only one particular application is shown in this paper, as an advanced hybrid simulation tool, this tool can be applied to many applications that involve large systems and require hybrid simulation to achieve high-fidelity results, including, but not limited to, detailed analysis of fault induced delayed voltage recovery events[8], transient stability analysis of power systems with HVDC [14] and/or FACTS, large wind or PV farm integration studies, dynamic simulation of distribution systems with a high penetration of DER [15], protection setting of distribution systems with a high penetration of DER. An older version of this tool has been applied to detailed analysis of fault induced delayed voltage recovery events on a large-scale planning case of the Western Interconnection of U.S., which consists of more than 15000 buses and 3000 generators, and a portion of the system with more than 200 buses (8 boundary buses) are modeled in detailed in the EMT part [8]. Thus, the scalability of the tool has been tested and proven in [8]. This tool has also been applied to detailed AC/DC power system dynamic simulations [14].

As this tool provides comprehensive modeling capabilities, and supports three-phase and three-sequence phasor domain modeling for the detailed and external systems, both balanced and unbalanced faults can be considered in either system. This overcomes many limitations associated with positive-sequence only TS and hybrid simulation approaches.

## VII. Conclusions

In this paper, an advanced EMT-TS hybrid simulation software package with the capabilities of comprehensive phasor-domain modeling and switching back to the TS simulation is developed. With this package, a whole simulation is divided into three stages, i.e., pre-fault TS simulation stage, faulted and post-fault hybrid simulation stage and post-disturbance TS simulation stage. To address the initialization issue of switching from EMT simulation to TS simulation, a three-phase phasor representation of the detailed system is simulated in parallel of the hybrid simulation. Furthermore, the critical discrete events captured in the EMT simulation are used to improve the accuracy of the dynamic simulation of the detailed system to address the simulation result discrepancy issue caused by the modeling differences between the EMT and the TS simulation. The test results show that, with the developed simulation switching method, the total computational time is significantly reduced compared to running the hybrid simulation for the whole simulation period, while a good accuracy is maintained.

## BIOGRAPHIES

**Qiuhua Huang** (S'14, M'16) received the B.E. and M.S. degree in electrical engineering from South China University of Technology, Guangzhou, China, in 2009 and 2012, respectively. He received his Ph.D. degree in electrical engineering from Arizona State University, Tempe, AZ, USA, in 2016. He is currently working as a research engineer at Pacific Northwest National Laboratory, Richland, WA, USA. He is broadly interested in power transmission and distribution system modeling, stability, simulation, and software development.

**Vijay Vittal** (S'78–F'97) received the B.E. degree in electrical engineering from the B.M.S. College of Engineering, Bangalore, India, in 1977, the M.Tech. degree from the Indian Institute of Technology, Kanpur, India, in 1979, and Ph.D. degree from Iowa State University, Ames, IA, USA, in 1982.

He is the Ira A. Fulton Chair Professor in the Department of Electrical, Computer and Energy Engineering at Arizona State University, Tempe, AZ, USA. He currently is the Director of the Power System Engineering Research Center (PSERC) Headquartered at Arizona State University.

Dr. Vittal is a member of the National Academy of Engineering.